\documentclass[pra,aps,eqsecnum,amsmath,amssymb,showpacs,showkeys]{revtex4}
\usepackage[mathscr]{eucal}
\newcommand{\hh}{\mathcal{H}}
\newcommand{\lsp}{{\mathcal{L}}_{+}}
\newcommand{\pen}{\openone}
\newcommand{\id}{{\rm{id}}}
\newcommand{\tr}{{\rm{tr}}}
\newcommand{\rqh}{{\rm{H}}_{q}^{(s)}}
\newcommand{\rqm}{{\rm{M}}_{q}^{(s)}}
\newcommand{\hg}{\rho^{Q}}
\newcommand{\fec}{{\Phi}^Q}
\newcommand{\rrg}{\rho^{Q'R'}}
\newcommand{\prq}{\psi^{QR}}
\newcommand{\ax}{\mathsf{X}}
\newcommand{\ay}{\mathsf{Y}}
\newcommand{\ip}{\mathsf{\Pi}}
\newcommand{\U}{\widetilde{\mathsf{U}}}
\newcommand{\lm}{\mathsf{\Lambda}}
\newcommand{\um}{\mathsf{U}}
\newcommand{\vm}{\mathsf{V}}
\newcommand{\am}{\mathsf{A}}
\newcommand{\bn}{\mathsf{B}}
\newcommand{\mm}{\mathsf{M}}
\newcommand{\np}{\mathsf{P}}
\newcommand{\ac}{\mathscr{A}}
\newcommand{\bc}{\mathscr{B}}
\newcommand{\pc}{\mathscr{C}}
\newcommand{\D}{\mathsf{D}}
\newcommand{\bah}{\bar{\rm{H}}}
\newcommand{\vrs}{\varsigma}
\newcommand{\cp}{\mathcal{C}}
\newtheorem{Thm}{Theorem}[section]
\newtheorem{Lem}[Thm]{Lemma}

\begin{document}

\preprint{}

\title{{\bf On unified-entropy characterization of quantum channels}}

\author{Alexey E. Rastegin}
 \affiliation{Department of Theoretical Physics, Irkutsk State University,
Gagarin Bv. 20, Irkutsk 664003, Russia}
 \email{rast@api.isu.ru}

\begin{abstract}
We consider properties of quantum channels with use of unified
entropies. Extremal unravelings of quantum channel with respect to
these entropies are examined. The concept of map entropy is
extended in terms of the unified entropies. The map
$(q,s)$-entropy is naturally defined as the unified
$(q,s)$-entropy of rescaled dynamical matrix of given quantum
channel. Inequalities of Fannes type are obtained for introduced
entropies in terms of both the trace and Frobenius norms of
difference between corresponding dynamical matrices. Additivity
properties of introduced map entropies are discussed. The known
inequality of Lindblad with the entropy exchange is generalized to
many of the unified entropies. For tensor product of a pair of
quantum channels, we derive two-side estimating of the output
entropy of a maximally entangled input state.
\end{abstract}



\maketitle

\section{Introduction}\label{intro}

The concept of entropy is one of most important notions in both
statistical physics and information theory. New applications of
this concept are connected with those advantages that can be
reached by using quantum resources to process and transmit the
information \cite{nielsen}. In addition to the Shannon and von
Neumann entropies, which are both fundamental, other entropic
measures were found to be useful. Among them, the R\'{e}nyi and
Tsallis $q$-entropic functionals are well known \cite{bengtsson}.
A general treatment of these and some other entropies in terms of
unified $(q,s)$-entropies was given in Ref. \cite{hey06}. For many
or even all values of parameters $q$ and $s$, the quantum unified
$(q,s)$-entropy enjoy features similar to properties of the
standard von Neumann entropy \cite{rastjst}. The unified entropies
was applied for treatment of quantum entanglement and monogamy
\cite{kims11}.

Entropic measures have widely been adopted in studying features of
quantum channels \cite{nielsen,bengtsson}. Different
characteristics are properly ensured by entropies of different
kinds and forms. The entropy exchange \cite{bsch96} and the map
entropy \cite{zb04} have been put to describe entanglement
transmission and decoherence induced by a quantum channel.
Additivity properties of map entropies with respect to the tensor
product of two channels are of interest \cite{rzf11}. For the
minimum output entropy, this question is considered to be even
more relevant \cite{shor04,horod10}. Together with the von Neumann
entropy, the quantum R\'{e}nyi entropy has been applied for these
purposes. In the present work, we treat characteristics of quantum
channels with use of the unified $(q,s)$-entropies.

The paper is organized as follows. The main definitions and the
notation are introduced in Section \ref{tsat}. In Section
\ref{exun}, we examine those channel unravelings that are extremal
with respect to the unified entropies. Continuity estimates of
Fannes type are derived for the map $(q,s)$-entropies in Section
\ref{cemes}. A distance between rescaled dynamical matrices are
quantified by means of both the trace and Frobenius norms. Section
\ref{bomap} is devoted to properties of the map $(q,s)$-entropies
with respect to the tensor product of two quantum channels. Using
an extension of Lindblad's inequality, we derive a two-sided
estimate on the output $(q,s)$-entropy for the tensor product of
two channels acting on maximally entangled input state. Section
\ref{concls} concludes the paper.

\section{Definitions and notation}\label{tsat}

Let $\hh$ be $d$-dimensional Hilbert space. We denote the space of
linear operators on $\hh$ by ${\mathcal{L}}(\hh)$ and the set of
positive semidefinite operators on $\hh$ by $\lsp(\hh)$. The
support of an operator is defined as the vector space orthogonal
to its kernel. A density operator $\rho\in\lsp(\hh)$ has unit
trace, i.e. $\tr(\rho)=1$. For $\ax,\ay\in{\mathcal{L}}(\hh)$, we
define the Hilbert--Schmidt inner product by \cite{watrous1}
\begin{equation}
\langle\ax{\,},\ay\rangle_{\rm{hs}}:=\tr(\ax^{\dagger}\ay)
\ . \label{hsdef}
\end{equation}
The Schatten norms, which form an important class of unitarily
invariant norms, are defined in terms of singular values. Recall
that singular values $\vrs_{j}(\ax)$ of operator $\ax$ are put as
eigenvalues of $|\ax|=\sqrt{\ax^{\dagger}\ax}$. For $q\geq1$, the
Schatten $q$-norm of $\ax\in{\mathcal{L}}({\mathcal{H}})$ is then
defined by \cite{watrous1,bhatia97}
\begin{equation}
\|{\mathsf{\ax}}\|_q=\left(\sum\nolimits_{j=1}^{d}\vrs_{j}({\mathsf{\ax}})^q\right)^{1/q}
\ . \label{shndf}
\end{equation}
This definition is closely related to the $q$-mean
${\mathfrak{M}}_{q}(x)=\left(\frac{1}{n}{\,}\sum_{j=1}^{n}x_{j}^{q}\right)^{1/q}$.
The properties of such means are extensively considered in the
book \cite{hardy}. The family (\ref{shndf}) includes the trace
norm $\|\ax\|_{1}$ for $q=1$, the Frobenius norm $\|\ax\|_{2}$ for
$q=2$, and the spectral norm
$\|\ax\|_{\infty}=\max\{\vrs_j(\ax):{\>}1\leq{j}\leq{d}\}$ for
$q=\infty$. The trace norm distance is one of most frequently used
distances. Its partitioned varieties are also be defined on the
base of Ky Fan's norms \cite{rast091,rast10}. In some respects,
however, we will prefer the Frobenius norm distance. Since the
Frobenius norm is induced by the Hilbert-Schmidt inner product,
i.e. $\|\ax\|_{2}^2=\langle\ax{\,},\ax\rangle_{\rm{hs}}$, this
distance is often named the Hilbert-Schmidt distance. Recently, it
has found to be fruitful in studying two-dimensional projections
of the set of mixed state \cite{dunkl11}.

The formalism of quantum operations provides a unified treatment
of possible state change in quantum theory
\cite{nielsen,watrous1}. Consider a linear map $\Phi$ that takes
elements of ${\mathcal{L}}(\hh)$ to elements of
${\mathcal{L}}(\hh')$, and also satisfies the condition of
complete positivity. Let $\id^R$ be the identity map on
$\mathcal{L}(\hh_R)$, where the space $\hh_R$ is assigned to
ancillary reference system. The complete positivity implies that
$\Phi\otimes\id^R$ transforms a positive operator into a positive
operator again for each dimension of the extended space. Such
linear maps are typically called ''quantum operations''
\cite{nielsen} or ''super-operators'' \cite{watrous1}. Each
completely positive map can be written in the operator-sum
representation. Namely, for any $\ax\in{\mathcal{L}}(\hh)$ we have
\begin{equation}
\Phi(\ax)=\sum\nolimits_{j} \am_{j}{\,}\ax{\,}\am_{j}^{\dagger}
\ , \label{oper3}
\end{equation}
where Kraus operators $\am_{j}$ map the input space $\hh$ to the
output space $\hh'$ \cite{nielsen,watrous1}. In general, the
normalization condition implies that
\begin{equation}
\sum\nolimits_{j}\am_{j}^{\dagger}\am_{j}\leq\pen
\ , \label{oper4}
\end{equation}
where $\pen$ is the identity operator on $\hh$. In most of
applications, the input and output spaces are the same. When the
physical process is deterministic, the equality in Eq.
(\ref{oper4}) holds and $\tr\bigl(\Phi(\rho)\bigr)=1$. In this
case, the map $\Phi$ is usually referred to as ''quantum channel''
\cite{nielsen}. Probabilistic operations, such as probabilistic
cloning \cite{duan2} or quantum state separation \cite{chefles1},
are of interest. But in the present paper, we will deal only with
deterministic quantum operations.

As an entropic measure, we will use the unified $(q,s)$-entropy
introduced in Ref. \cite{hey06} and further studied in Ref.
\cite{rastjst}. The $(q,s)$-entropies form a family of
two-parameter entropic functionals continuous with respect to both
the parameters \cite{hey06}. Many generalized entropies including
the R\'{e}nyi and Tsallis ones are contained in this family. The
quantum unified $(q,s)$-entropy of density operator $\rho$ is
defined as \cite{hey06}
\begin{equation}
\rqh(\rho):=\frac{1}{(1-q){\,}s}{\>}
\Bigl\{\bigl[\tr(\rho^{q})\bigr]^s-1\Bigr\}
\label{uqndef}
\end{equation}
for $q>0$, $q\neq1$ and $s\neq0$. For $q=1$, this entropy is
defined as the von Neumann entropy
${\rm{H}}_1(\rho)=-\tr(\rho\ln\rho)$. For $s=1$, we obtain the
quantum Tsallis $q$-entropy
\begin{equation}
{\rm{T}}_{q}(\rho):=\frac{1}{1-q}{\>}\tr\bigl(\rho^{q}-\rho\bigr)=\tr\bigl(\eta_q(\rho)\bigr)
\ , \label{qtsdf}
\end{equation}
where $\eta_q(x):=\bigl(x^q-x\bigr)\big/(1-q)=-x^{q}\ln_{q}x$ in
terms of $q$-logarithm $\ln_{q}x=\bigl(x^{1-q}-1\bigr)\big/(1-q)$.
In the limit $s\to0$, the definition (\ref{uqndef}) leads to the
quantum R\'{e}nyi $q$-entropy $\rho$ defined as
\begin{equation}
{\rm{R}}_{q}(\rho):=\frac{1}{1-q}{\>}\ln\bigl[\tr(\rho^{q})\bigr]
\ . \label{qredf}
\end{equation}
For $q=1$, both the expressions (\ref{qtsdf}) and (\ref{qredf})
recover the von Neumann entropy. The classical entropies can all
be obtained by replacing the traces with the proper sums over a
probability distribution. Let $X$ be random variable, taking $m$
possible values with probabilities $p_X(i)$ ($i=1,\ldots,m)$. Its
$(q,s)$-entropy is
\begin{equation}
H_{q}^{(s)}(X):=\frac{1}{(1-q){\,}s}{\,}\left[\left(\sum\nolimits_{i=1}^m
p_X(i)^{q}\right)^{{\!}s}-1\right]
\label{unfdef}
\end{equation}
for $q>0$, $q\neq1$ and $s\neq0$, and the Shannon entropy
$H_{1}(X)=-\sum_{i}p_X(i)\ln{p}_X(i)$ for $q=1$. For $s=1$, the
right-hand side of Eq. (\ref{unfdef}) gives the classical Tsallis
$q$-entropy $T_{q}(X)$. In the limit $s\to0$, the unified
$(q,s)$-entropy recovers the Renyi $q$-entropy
\begin{equation}
R_{q}(X)=\frac{1}{1-q}{\>}\ln\left(\sum\nolimits_{i=1}^m p_X(i)^{q}\right)
\ . \label{rndef}
\end{equation}
Two specific entropic measures will be adopted through the paper.
The first is the $(q,s)$-entropy exchange. For its description, we
have to separate explicitly the principal system $Q$ from an
imagined reference system $R$ and an environment $E$. The Hilbert
spaces are denoted by $\hh_Q$, $\hh_R$, and $\hh_E$, respectively.
Under the action of quantum channel $\fec$, the initial state
$\hg$ of system $Q$ is mapped into $\fec(\hg)$. To see the
entanglement transmission, we consider a purification
$|\prq\rangle\in\hh_Q\otimes\hh_R$, which is transformed into the
final state
\begin{equation}
\rrg=\fec\otimes\id^R\bigl(|\prq\rangle\langle\prq|\bigr)
\label{strqe}
\end{equation}
of the system $QR$. The system $R$ itself is not altered, i.e.
$\tr_Q\bigl(\rrg\bigr)=\tr_Q\left(|\prq\rangle\langle\prq|\right)$.
Putting an environment $E$, we can reexpress the quantum channel
$\fec$ in terms of unitary operator $\U$ on $\hh_E\otimes\hh_Q$ as
\begin{equation}
\fec(\hg)=\tr_E\left(\U(|e_0\rangle\langle e_0|\otimes\hg)\U^{\dagger}\right)
\ . \label{quopc}
\end{equation}
Since the final state
$(\U\otimes\pen^R)|e_0\rangle\otimes|\prq\rangle$ of the triple
system $EQR$ is obviously pure, the final density operators
$\rho^{E'}\in\lsp(\hh_E)$ and $\rrg\in\lsp(\hh_Q\otimes\hh_R)$
have the same non-zero eigenvalues. We define the $(q,s)$-entropy
exchange as
\begin{equation}
\bah_{q}^{(s)}\left(\hg,\fec\right):=\rqh(\rrg)=\rqh(\rho^{E'})
\ . \label{qseedf}
\end{equation}
This quantity characterizes an amount of $(q,s)$-entropy
introduced by the quantum channel $\fec$ into an initially pure
environment $E$. The definition (\ref{qseedf}) is a direct
extension of the Schumacher entropy exchange \cite{bsch96} to the
considered entropic measure. It can be shown that the right-hand
side of Eq. (\ref{qseedf}) depends only on the initial state $\hg$
of the principal system and the quantum channel $\Phi^Q$ (for
details, see subsection 12.4.1 in \cite{nielsen}). So further we
can left out the superscript $Q$ of the principal system and
merely write $\bah_{q}^{(s)}(\rho,\Phi)$.

Another specific entropic measure is the map $(q,s)$-entropy. It
is defined within the Jamio{\l}kowski--Choi representation of $\Phi$
\cite{jam72,choi75}. Let $\hh_Q=\hh_R=\hh$ and $\{|\nu\rangle\}$
be an orthonormal basis in $\hh$. To this basis we assign the
normalized pure state
\begin{equation}
|\phi_{+}\rangle:=\frac{1}{\sqrt{d}}{\,}\sum_{\nu=1}^{d}
{|\nu\rangle\otimes|\nu\rangle} \ , \label{phip}
\end{equation}
where $d$ is the dimensionality of $\hh$. We then put the special
operator
\begin{equation}
\sigma(\Phi):=\Phi\otimes\id\bigl(|\phi_{+}\rangle\langle\phi_{+}|\bigr)
\ , \label{sigmat}
\end{equation}
acting on the doubled space $\hh^{\otimes2}$. The matrix
$\D(\Phi)=d{\,}\sigma(\Phi)$ is called ''dynamical matrix'' or
''Choi matrix'' \cite{choi75}. For each
$\ax\in{\mathcal{L}}(\hh)$, the action of super-operator $\Phi$
can be recovered from $\D(\Phi)$ by means of the relation
\cite{watrous1}
\begin{equation}
\Phi(\ax)=\tr_R\left(\D(\Phi)(\pen\otimes\ax^{T})\right)
\ , \label{chjis}
\end{equation}
where $\ax^{T}$ denotes the transpose operator to $\ax$. The map
$\Phi$ is completely positive, whenever the matrix $\D(\Phi)$ is
positive. The condition $\tr_R\bigl(\D(\Phi)\bigr)=\pen$ is
equivalent to that the map $\Phi$ is trace-preserving and the
rescaled matrix $\sigma(\Phi)$ is of unit trace. For given channel
$\Phi$, we define the map $(q,s)$-entropy by
\begin{equation}
\rqm(\Phi):=\rqh\bigl(\sigma(\Phi)\bigr)
\ . \label{mapqs}
\end{equation}
This is an extension of the standard map entropy introduced in
\cite{zb04} and further examined in \cite{rzf11}. The map entropy
is used to characterize the decoherent behaviour of given channel.

\section{Extremal unravelings of a quantum channel}\label{exun}

In this section, we study extremality of unravelings of a quantum
channel with respect to unified entropies. Recall that
representations of the form (\ref{oper3}) are not unique
\cite{watrous1}. For given map $\Phi$, there are many sets
$\ac=\{\am_{j}\}$ that enjoy Eq. (\ref{oper3}). In the paper
\cite{ilichev03}, each concrete set $\ac=\{\am_j\}$ resulting in Eq.
(\ref{oper3}) is named an ''unraveling'' of the map $\Phi$. This
terminology is due to Carmichael \cite{carm} who introduced this
word for a representation of the master equation (for a review of
this topic, see Ref. \cite{wm10}). Following the method of Ref.
\cite{rast104}, we introduce the matrix
\begin{equation}
\ip(\ac|\rho):=[[\langle\am_i\sqrt{\rho}{\,},\am_j\sqrt{\rho}\rangle_{\rm{hs}}]]=
[[\tr(\am_i^{\dagger}\am_j\rho)]]
\ , \label{pimdef}
\end{equation}
for given density operator $\rho$ and channel unraveling
$\ac=\{\am_i\}$. The diagonal element
$p_i=\tr(\am_i^{\dagger}\am_i\rho)$ is clearly positive and gives
the $i$th effect probability. Then the entropy $\rqh(\ac|\rho)$ is
defined by Eq. (\ref{unfdef}). It is well-known that two operator-sum
representations of the same completely positive map are related as
\begin{equation}
\bn_{i}=\sum\nolimits_{j} \am_j{\,}u_{ji}
\ , \label{eqvun}
\end{equation}
where the matrix $\um=[[u_{ij}]]$ is unitary \cite{nielsen}. This
statement can be obtained on the base of the ensemble
classification theorem proved in Ref. \cite{hugston}. If the two
sets $\ac=\{\am_j\}$ and $\bc=\{\bn_i\}$ fulfill Eq. (\ref{eqvun}),
then we have
\begin{equation}
\langle\bn_i\sqrt{\rho}{\,},\bn_k\sqrt{\rho}\rangle_{\rm{hs}}=
\sum\nolimits_{jl} u_{ji}^{*}{\,}u_{lk}{\,}
\langle\am_j\sqrt{\rho}{\,},\am_l\sqrt{\rho}\rangle_{\rm{hs}}
\ , \label{pinner}
\end{equation}
or merely $\ip(\bc|\rho)=\um^{\dagger}{\,}\ip(\ac|\rho){\,}\um$.
In other words, the matrices $\ip(\ac|\rho)$ and $\ip(\bc|\rho)$
are unitarily similar \cite{rast104}. By Hermiticity, all such
matrices assigned to the same channel are unitarily similar to a
unique (up to permutations) diagonal matrix
$\lm=\rm{diag}(\lambda_1,\lambda_2,\ldots)$, where the
$\lambda_i$'s are the eigenvalues of each of these matrices. Hence
any $\ip(\ac|\rho)$ is positive semidefinite. For given unraveling
$\ac=\{\am_j\}$, we obtain the concrete matrix $\ip(\ac|\rho)$ and
diagonalize it through a unitary transformation
$\vm^{\dagger}{\,}\ip(\ac|\rho){\,}\vm=\lm$. Let us define a
specific unraveling $\ac_{\rho}^{(ex)}$ related to given $\ac$ as
\begin{equation}
\am_{i}^{(ex)}=\sum\nolimits_j \am_j{\,} v_{ji}
\ , \label{excalc}
\end{equation}
where the unitary matrix $\vm=[[v_{ij}]]$ diagonalizes
$\ip(\ac|\rho)$. We shall show that the unraveling (\ref{excalc})
enjoys the extremality property with respect to almost all the
$(q,s)$-entropies.

\begin{Thm}\label{extm}
Let $\rho$ be density operator on $\hh$, and $\ac$ an unraveling
of quantum channel. For all $q>0$ and $s\neq0$, there holds
\begin{eqnarray}
H_{q}^{(s)}(\ac_{\rho}^{(ex)}|\rho)\leq{H}_{q}^{(s)}(\ac|\rho)
 \label{unex}
\end{eqnarray}
where the extremal unraveling $\ac_{\rho}^{(ex)}$ is defined by
Eq. (\ref{excalc}).
\end{Thm}

{\bf Proof.} We first suppose that $q\neq1$. By construction, we
have $\ip(\ac_{\rho}^{(ex)}|\rho)=\lm$ with the probabilities
$\lambda_j$ of effects. Due to
$\ip(\ac|\rho)=\vm{\,}\lm{\,}\vm^{\dagger}$, the probabilities
$p_i=\tr(\am_i^{\dagger}\am_i\rho)$ of different effects of $\ac$
is related to the $\lambda_j$'s by
\begin{equation}
p_i=\sum\nolimits_j v_{ij}{\,}\lambda_j{\,}v_{ij}^{*}
=\sum\nolimits_{j} w_{ij}{\,}\lambda_j \ . \label{pipij}
\end{equation}
Here numbers $w_{ij}=v_{ij}{\,}v_{ij}^{*}$ are elements of
unistochastic matrix, whence $\sum\nolimits_{i}w_{ij}=1$ for all
$j$ and $\sum\nolimits_{j}w_{ij}=1$ for all $i$. The function
$x\mapsto{x}^{q}$ is concave for $0<q<1$ and convex for $1<q$.
Applying Jensen's inequality to this function, we obtain
\begin{equation}
\sum\nolimits_i p_i^{q}=
\sum\nolimits_i\left(\sum\nolimits_{j} w_{ij}{\,}\lambda_j\right)^{q}
{\ }\left\{
\begin{array}{cc}
\geq, & 0<q<1 \\
\leq, & 1<q
\end{array}
\right\}
{\ }\sum\nolimits_i \sum\nolimits_{j} w_{ij}{\,}\lambda_j^{q}
=\sum\nolimits_j \lambda_j^{q}
\label{agal1}
\end{equation}
in view of the above unistochasticity. The function $y\mapsto
y^s/s$ monotonically increases for all $s\neq0$, whence
\begin{equation}
\frac{1}{s}{\>}\left(\sum\nolimits_{i} p_i^{q}\right)^{{\!}s}
{\ }\left\{
\begin{array}{cc}
\geq, & 0<q<1 \\
\leq, & 1<q
\end{array}
\right\} {\ }\frac{1}{s}{\>}\left(\sum\nolimits_{j}
\lambda_j^{q}\right)^{{\!}s}
\ . \label{agal2}
\end{equation}
Since the term $(1-q)$ is positive for $q<1$ and negative for
$1<q$, the relations (\ref{agal2}) are merely combined as
\begin{equation}
\frac{1}{(1-q){\,}s}{\>}\left(\sum\nolimits_i p_i^{q}\right)^{{\!}s}
\geq\frac{1}{(1-q){\,}s}{\>}\left(\sum\nolimits_j \lambda_j^{q}\right)^{{\!}s}
\ . \label{agal3}
\end{equation}
Due to the definition (\ref{unfdef}), the inequality (\ref{agal3})
provides (\ref{unex}). In the case $q=1$, we deal with the Shannon
entropy. Applying Jensen's inequality to the concave function
$x\mapsto-x\ln{x}$ completes the proof of this case.
$\blacksquare$

For prescribed state $\rho$ and given unraveling of quantum
channel all the unified $(q,s)$-entropies with parameter $s\neq0$
are minimized by the extremal unraveling, which is built in line
with Eq. (\ref{excalc}). For the Shannon entropy, a question with
''minimal'' unraveling was considered in Ref. \cite{breslin} and
later in Ref. \cite{ilichev03}. It was noted in Ref.
\cite{rast104} that diagonalizing the matrix $\ip(\ac|\rho)$ is
formally equivalent to the extreme condition of Ref.
\cite{ilichev03}. The latter condition has been obtained by
calculation with Lagrange's multipliers which is rather local in
spirit. These reasons are complemented by the above proof based
purely on the concavity property. As it was shown in Ref.
\cite{rast104}, the R\'{e}nyi entropies enjoy the extremality
property with unraveling (\ref{excalc}) only for order
$q\in(0;1)$. In this regard, the R\'{e}nyi entropies differ from
other considered entropies.

A quantum state is characterized by the probabilities of the
outcomes of every conceivable test \cite{peresq}. So the
measurements are quantum operations of special conceptual
interest. A general measurement is described by the set
$\{\mm_i\}$ of measurement operators \cite{nielsen}. Suppose that
$\rho$ is density operator of the system right before the
measurement. Separate terms of the sum
\begin{equation}
\sum\nolimits_i \mm_i{\,}\rho{\,}\mm_i^{\dagger}
\label{mrom}
\end{equation}
are related to different outcomes of the measurement. Since the
probabilities $\tr\bigl(\mm_i^{\dagger}\mm_i\rho\bigr)$ are summed
to one, the measurement operators enjoy the equality in Eq.
(\ref{oper4}). A standard measurement is described by the set
$\{\np_j\}$ of mutually orthogonal projectors. As a rule,
projective measurements are easier to realize experimentally. One
of basic properties of the von Neumann entropy is that it cannot
be decreased by a projective measurement (see, e.g., theorem 11.9
in \cite{nielsen}). For given measurement $\{\np_j\}$, the output
density operator is expressed as \cite{nielsen}
\begin{equation}
\cp(\rho)=\sum\nolimits_j \np_j{\,}\rho{\,}\np_j
\ . \label{afms}
\end{equation}
In matrix analysis, an operation of such a kind is referred to as
''pinching'' (for details, see section IV.2 of Ref.
\cite{bhatia97}). The non-decreasing of the von Neumann entropy is
posed as ${\rm{H}}_1(\rho)\leq{\rm{H}}_1\bigl(\cp(\rho)\bigr)$. As
it is shown in Ref. \cite{rastjst}, all the unified entropies are
non-decreasing under projective measurements, namely
\begin{equation}
\rqh(\rho)\leq\rqh\bigl(\cp(\rho)\bigr)
\ . \label{uendn}
\end{equation}
Hence we can see that
$\rqh(\rho)\leq{H}_{q}^{(s)}(\pc_{\rho}^{(ex)}|\rho)$, where
$\pc_{\rho}^{(ex)}$ denotes extremal unraveling of the projective
measurement. In other words, the entropy $\rqh(\rho)$ is a lower
bound for the extremal unraveling entropy
$H_{q}^{(s)}(\pc_{\rho}^{(ex)}|\rho)$. The latter inequality is
always saturated, when the measurement is carried out in the basis
of the eigenstates of $\rho$. We shall extend this treatment to
other quantum channels. Such an extension takes place, when the
Kraus operators of extremal unraveling fulfill a certain
condition.

\begin{Thm}\label{mext}
Suppose that the Kraus operators $\am_i^{(ex)}$ of
extremal unraveling of a quantum channel satisfy
\begin{equation}
\tr\bigl(\am_i^{(ex)\dagger}\am_i^{(ex)}\bigr)=1
\label{cond}
\end{equation}
for all values of index $i$. For $0<q$ and $s\neq0$ as well as for
$0<q<1$ and $s=0$, we then have
\begin{equation}
\rqh(\rho)\leq{H}_{q}^{(s)}(\ac_{\rho}^{(ex)}|\rho)
\ . \label{haer}
\end{equation}
\end{Thm}

{\bf Proof.} We first suppose that $q\neq1$. By definition, the
matrix $\ip(\ac_{\rho}^{(ex)}|\rho)$ is diagonal. Calculating the
trace in the eigenbasis $\{|j\rangle\}$ of $\rho$, we rewrite
diagonal elements as
\begin{equation}
\lambda_i=\tr\bigl(\am_i^{(ex)\dagger}\am_i^{(ex)}\rho\bigr)=\sum\nolimits_j
\vrs_j\langle{j}|\am_i^{(ex)\dagger}\am_i^{(ex)}|j\rangle
\ , \label{diel}
\end{equation}
where the $\vrs_j$'s are eigenvalues of $\rho$. The numbers
$t_{ij}=\langle{j}|\am_i^{(ex)\dagger}\am_i^{(ex)}|j\rangle$ form
a double-stochastic matrix due to
\begin{equation}
\sum\nolimits_j{t_{ij}}=\tr\bigl(\am_i^{(ex)\dagger}\am_i^{(ex)}\bigr)=1
\ , \qquad
\sum\nolimits_i{t_{ij}}=\langle{j}|j\rangle=1
\ . \label{cond1}
\end{equation}
Here the first follows from Eq. (\ref{cond}), the second follows
from the equality in Eq. (\ref{oper4}). Combining both the
relations of Eq. (\ref{cond1}) with the Jensen inequality, we
obtain
\begin{equation}
\sum\nolimits_i\lambda_i^{q}=
\sum\nolimits_i\left(\sum\nolimits_{j} t_{ij}{\,}\vrs_j\right)^{q}
{\ }\left\{
\begin{array}{cc}
\geq, & 0<q<1 \\
\leq, & 1<q
\end{array}
\right\}
{\ }\sum\nolimits_i \sum\nolimits_{j} t_{ij}{\,}\vrs_j^{q}
=\sum\nolimits_j \vrs_j^{q}
\ . \label{agal11}
\end{equation}
As the function $y\mapsto y^s/s$ monotonically increases for
$s\neq0$, these inequalities lead to
\begin{equation}
\frac{1}{s}{\>}\left(\sum\nolimits_i\lambda_i^{q}\right)^{{\!}s}
{\ }\left\{
\begin{array}{cc}
\geq, & 0<q<1 \\
\leq, & 1<q
\end{array}
\right\} {\ }\frac{1}{s}{\>}\left(\sum\nolimits_j
\vrs_j^{q}\right)^{{\!}s}
\ . \label{agal21}
\end{equation}
Multiplying Eq. (\ref{agal21}) by the factor $(1-q)^{-1}$, which
is positive for $0<q<1$ and negative for $1<q$, we complete the
proof for $q\neq1$ and $s\neq0$. In the case $q=1$, when the von
Neumann and Shannon entropies are dealt, we merely combine Eq.
(\ref{cond1}) with the concavity of the function
$x\mapsto-x\ln{x}$. When $s=0$, we apply the relations
\begin{equation}
{\rm{R}}_{q}(\rho)=(1-q)^{-1}\ln\Bigl(1+(1-q){\rm{T}}_{q}(\rho)\Bigr)
\ , \qquad
R_{q}(\ac_{\rho}^{(ex)}|\rho)=(1-q)^{-1}\ln\Bigl(1+(1-q)T_{q}(\ac_{\rho}^{(ex)}|\rho)\Bigr)
\ , \label{rentsa}
\end{equation}
which at once follow from the definitions of the R\'{e}nyi and
Tsallis $q$-entropies. Since the inequality (\ref{haer}) holds for
the Tsallis case ($s=1$) and the function
$y\mapsto(1-q)^{-1}\ln\bigl(1+(1-q){\,}y\bigr)$ is increasing for
$q<1$, the R\'{e}nyi $q$-entropies also enjoy Eq. (\ref{haer}) for
$0<q<1$. $\blacksquare$

Thus, if the condition (\ref{cond}) holds then the extremal
unraveling entropy ${H}_{q}^{(s)}(\ac_{\rho}^{(ex)}|\rho)$ is
bounded from below by the entropy of the input state $\rqh(\rho)$.
This property takes place for all unified $(q,s)$-entropies,
except for the R\'{e}nyi $q$-entropies of order $q>1$. Here we
again see some distinctions of the R\'{e}nyi entropies from the
rest entropies considered. The inequality (\ref{haer}) may also be
regarded as an estimate on the entropy of input state of a
channel, when the extremal unraveling entropy is known (exactly or
approximately) from other reasons.

\section{Continuity estimates on the map $(q,s)$-entropies}\label{cemes}

One of essential properties of the von Neumann entropy is its
continuity firstly stated by Fannes \cite{fannes}. Fannes'
inequality has been generalized to the Tsallis entropy
\cite{yanagi,zhang} and its partial sums \cite{rast1023}.
Continuity estimates of Fannes type have also been derived for the
quantum conditional entropy \cite{alf04} as well as for the
standard quantum relative entropy \cite{AE05} and its
$q$-extension \cite{rastjmp}. Below we will consider inequalities
of Fannes type for the map $(q,s)$-entropies. Continuity estimates
on the unified $(q,s)$-entropies were derived in Ref.
\cite{rastjst}. They are based on the known inequalities with the
Tsallis entropies \cite{yanagi,zhang}. Combining the estimates of
Ref. \cite{rastjst} with the definition (\ref{mapqs}), we obtain
the following. Let $\Phi$, $\Psi$ be quantum channels on
$d$-dimensional states. For the parameter range
\begin{equation}
\bigl\{(q,s):{\>}0<q<1,{\>}s\in(-\infty;-1]\cup[0;+1]\bigr\}
\label{pdm1}
\end{equation}
under the condition
$\|\sigma(\Phi)-\sigma(\Psi)\|_{1}=2t\leq{q}^{1/(1-q)}$, there holds
\begin{equation}
\bigl|\rqm(\Phi)-\rqm(\Psi)\bigr|
\leq(2t)^{q}\ln_{q}d+\eta_{q}(2t)
\ . \label{mesfk1}
\end{equation}
Define the factor $\varkappa_s=d^{2(q-1)}$ for $s\in[-1;0]$ and
$\varkappa_s=1$ for $s\in[+1;+\infty)$. For the parameter range
\begin{equation}
\bigl\{(q,s):{\>}1<q,{\>}s\in[-1;0]\cup[+1;+\infty)\bigr\}
\label{pdp1}
\end{equation}
under the condition
$(1/2)\|\sigma(\Phi)-\sigma(\Psi)\|_{1}=t\leq(d-1)/d$, there holds
\begin{equation}
\bigl|\rqm(\Phi)-\rqm(\Psi)\bigr|
\leq\varkappa_s\bigl[t^{q}\ln_{q}(d-1)+T_{q}(t,1-t)\bigr]
\ . \label{meszh1}
\end{equation}
Here $T_{q}(t,1-t)$ is the binary Tsallis entropy. The validity
ranges $0\leq2t\leq{q}^{1/(1-q)}$ for (\ref{mesfk1}) and
$0\leq{t}\leq(d-1)/d$ for (\ref{meszh1}) are essential and
obtained as intervals of non-decreasing of the corresponding
right-hand sides.

In principle, the continuity property is established by means of
Eqs. (\ref{mesfk1}) and (\ref{meszh1}). At the same time, the norm
$\|\sigma(\Phi)-\sigma(\Psi)\|_{1}$ is sufficiently difficult for
calculation in general form. The Frobenius norm distance
$\|\sigma(\Phi)-\sigma(\Psi)\|_{2}$ is easier to estimate. By Eq.
(\ref{sigmat}), there holds
\begin{equation}
|\sigma(\Phi)-\sigma(\Psi)|^2=\frac{1}{d^2}{\>}\sum_{\mu\nu\xi}
\bigl\{\Phi(|\mu\rangle\langle\nu|)-\Psi(|\mu\rangle\langle\nu|)\bigr\}
\bigl\{\Phi(|\nu\rangle\langle\xi|)-\Psi(|\nu\rangle\langle\xi|)\bigr\}
\otimes|\mu\rangle\langle\xi|
\ . \label{mnxi0}
\end{equation}
Hence, in view of
$\tr\bigl(\ax\otimes\ay\bigr)=\tr(\ax){\,}\tr(\ay)$ and the
linearity of the trace, we obtain
\begin{equation}
\tr\left(|\sigma(\Phi)-\sigma(\Psi)|^2\right)=\frac{1}{d^2}{\>}\sum_{\mu\nu}
{\tr\left(|\Phi(|\mu\rangle\langle\nu|)-\Psi(|\mu\rangle\langle\nu|)|^2\right)}
=\frac{1}{d^2}{\>}\sum_{\mu\nu}{\bigl\|\Phi(|\mu\rangle\langle\nu|)-\Psi(|\mu\rangle\langle\nu|)\bigr\|_{2}^2}
\ . \label{mnxi1}
\end{equation}
In other words, the Frobenius norm distance
$\|\sigma(\Phi)-\sigma(\Psi)\|_{2}$ is expressed as the 2-mean
\begin{equation}
\|\sigma(\Phi)-\sigma(\Psi)\|_{2}={\mathfrak{M}}_2\bigl(\|(\Phi-\Psi)(|\mu\rangle\langle\nu|)\|_2\bigr)
\label{mp2m}
\end{equation}
of the particular distances
$\|(\Phi-\Psi)(|\mu\rangle\langle\nu|)\|_2$ taken with equal
weights. Other questions, in which the Frobenius norm distance is
very useful, are treated in Ref. \cite{dunkl11}. So, it is of
interest to pose continuity estimates on the map entropies in
terms of the distance (\ref{mp2m}). On the other hand, the
inequalities of Fannes type are naturally formulated in terms of
the trace norm distance. Hence we are interested in relations
between the trace and Frobenius norms. This issue is considered in
Appendix \ref{srsn}. Taking $q=2$ in (\ref{n1npq}), we obtain
$\|\ax\|_{1}\leq\sqrt{d}{\>}\|\ax\|_{2}$ for each
$\ax\in{\mathcal{L}}(\hh)$.

In the intervals of non-decreasing right-hand sides, the upper
bounds (\ref{mesfk1}) and (\ref{meszh1}) are recast with larger
$\sqrt{d}{\>}\|\sigma(\Phi)-\sigma(\Psi)\|_{2}$ instead of
$\|\sigma(\Phi)-\sigma(\Psi)\|_{1}$ as follows. For the parameter
range (\ref{pdm1}), there holds
\begin{equation}
\bigl|\rqm(\Phi)-\rqm(\Psi)\bigr|
\leq{d}^{q/2}(2\tau)^{q}\ln_{q}d+\eta_{q}\bigl(\sqrt{d}{\,}2\tau\bigr)
\ , \label{mesfk2}
\end{equation}
provided that
$\|\sigma(\Phi)-\sigma(\Psi)\|_{2}=2\tau\leq{q}^{1/(1-q)}d^{-1/2}$.
For the parameter range (\ref{pdp1}), there holds
\begin{equation}
\bigl|\rqm(\Phi)-\rqm(\Psi)\bigr|
\leq\varkappa_s\left[{d}^{q/2}\tau^{q}\ln_{q}(d-1)
+T_{q}\bigl(\sqrt{d}{\,}\tau,1-\sqrt{d}{\,}\tau\bigr)\right]
\ , \label{meszh2}
\end{equation}
provided that
$(1/2)\|\sigma(\Phi)-\sigma(\Psi)\|_{2}=\tau\leq(d-1){\,}d^{-3/2}$.
So, we have arrived at upper bounds in terms of the Frobenius norm
distance between the two rescaled dynamical matrices. However, the
validity ranges of Eqs. (\ref{mesfk2}) and (\ref{meszh2}),
particularly the former, seem to be too restrictive for
sufficiently large $d$. Since the operator
$\sigma(\Phi)-\sigma(\Psi)$ is traceless Hermitian and
$\tr\bigl(\sigma(\Phi)\bigr)=\tr\bigl(\sigma(\Psi)\bigr)=1$, the
trace and Frobenius norms obey
$\|\sigma(\Phi)-\sigma(\Psi)\|_{1}\leq2$ and
$\|\sigma(\Phi)-\sigma(\Psi)\|_{2}\leq\sqrt{2}$, respectively.
Below, we will derive upper bounds that hold for all
$\|\sigma(\Phi)-\sigma(\Psi)\|_{2}\leq\sqrt{2}$.

Due to Csisz\'{a}r, the classical Fano inequality leads to an
improvement of the original Fannes' bound (see theorem 3.8 and its
proof in Ref. \cite{petz08}). Developing this point was for the
Tsallis entropies \cite{rast1101}, we obtain the inequality
\begin{equation}
\bigl|{\rm{T}}_{q}(\rho)-{\rm{T}}_{q}(\omega)\bigr|\leq
{t}^{q}\ln_{q}\bigl[d(d-1)\bigr]+
\frac{t^{q}-qt}{1-q}=\frac{\bigl(d(d-1)\bigr)^{1-q}t^{q}-qt}{1-q}
\label{tql1}
\end{equation}
for $0<q<1$ and all $(1/2)\|\rho-\omega\|_{1}=t\leq1$, and the
inequality
\begin{equation}
\bigl|{\rm{T}}_{q}(\rho)-{\rm{T}}_{q}(\sigma)\bigr|\leq
t^{q}\ln_{q}(d-1)+T_{q}(t,1-t)
\label{tqg1}
\end{equation}
for $1<q$ and $(1/2)\|\rho-\omega\|_{1}=t\leq(d-1)/d$. The second
is exactly those bound that was derived immediately in Ref.
\cite{zhang}. In comparison with Eq. (\ref{mesfk1}), the upper
bound (\ref{tql1}) is valid for all acceptable values of the norm
$\|\rho-\omega\|_{1}$, including its maximal value $2$. We shall
now modify Eq. (\ref{tqg1}) in this regard. The bound (\ref{tqg1})
was derived in Ref. \cite{rast1101} with use of the relation
\begin{equation}
T_{q}(X)-T_{q}(Y)\leq{T}_{q}(X|Y)\leq{P}_e^{{\,}q}\ln_{q}(m-1)+{T}_{q}(P_e,1-P_e)
\ . \label{infn}
\end{equation}
Here random variables $X$ and $Y$ take the same $m$ possible
values, $T_{q}(X|Y)$ is the conditional $q$-entropy, and $P_e$ is
the probability of error, i.e. the probability of that $X\neq{Y}$.
The right-hand side of Eq. (\ref{infn}) is an extension of the
classical Fano inequality to the conditional Tsallis entropy (for
its properties, see Ref. \cite{sf06}). We rewrite the right-hand
side of Eq. (\ref{infn}) in the form
\begin{equation}
\frac{1-(1-P_e)^q-(m-1)^{1-q}P_e^{{\,}q}}{q-1}\leq
\frac{q{P_e}-(m-1)^{1-q}P_e^{{\,}q}}{q-1}
\ , \label{pinq}
\end{equation}
which follows from the inequality
$1-(1-P_e)^q=\int_{0}^{P_e}{q(1-x)^{q-1}dx}\leq\int_{0}^{P_e}{q{\,}dx}=q{P_e}$
in view of $q>1$. As function of $P_e$, the left-hand side of Eq.
(\ref{pinq}) monotonically increases up to the point
$P_e=(m-1)/m$, where the derivative is zero. But the derivative of
right-hand side of Eq. (\ref{pinq}) vanishes at the point
$P_e=m-1$. So the interval of non-decreasing becomes wider, though
the bound with the right-hand side of Eq. (\ref{pinq}) will
somewhat weaker. Setting $m=d$, we consider the probabilities
of $X$ and $Y$ as eigenvalues of $\rho$ and $\omega$,
respectively. The joint probability mass function of $X$ and $Y$
can be built in such a way that
$P_e=(1/2)\sum_i|p_X(i)-p_Y(i)|\leq(1/2)\|\rho-\omega\|_{1}$ (the
equality on the left is a part of the coupling inequality
\cite{lindvall}, the inequality on the right follows from lemma
11.1 of Ref. \cite{petz08}). Combining this with Eq. (\ref{infn})
leads to the inequality of Fannes type in the form
\begin{equation}
\bigl|{\rm{T}}_{q}(\rho)-{\rm{T}}_{q}(\omega)\bigr|\leq
\frac{q{t}-(d-1)^{1-q}t^{q}}{q-1}
\ , \label{wfin}
\end{equation}
where $t=(1/2)\|\rho-\omega\|_{1}$ may take all possible values
from the range $0\leq{t}\leq1$. Replacing $t$ with larger
$\sqrt{d}{\,}\tau$, we get the upper bound
\begin{equation}
\bigl|{\rm{T}}_{q}(\rho)-{\rm{T}}_{q}(\omega)\bigr|\leq
\frac{q\sqrt{d}{\,}\tau-(d-1)^{1-q}d^{q/2}\tau^{q}}{q-1}
\ , \label{wfin1}
\end{equation}
where $\tau=(1/2)\|\rho-\omega\|_{2}$. The above replacing is
valid under the constraint $\sqrt{d}{\,}\tau\leq{d}-1$, i.e.
$\|\rho-\omega\|_{2}\leq2(d-1)/\sqrt{d}$. The latter always holds,
since $\|\rho-\omega\|_{2}\leq\sqrt{2}$ for two density matrices
and $\sqrt{d/2}\leq{d}-1$ for all $d\geq2$. In other words, the upper
bound (\ref{wfin1}) is dealt for all possible values of the
Frobenius norm distance between states (and $1<q$). In the same
manner, from Eq. (\ref{tql1}) we obtain
\begin{equation}
\bigl|{\rm{T}}_{q}(\rho)-{\rm{T}}_{q}(\omega)\bigr|\leq
\frac{d^{1-q/2}(d-1)^{1-q}\tau^{q}-q\sqrt{d}{\,}\tau}{1-q}
\ . \label{tql2}
\end{equation}
The right-hand side of Eq. (\ref{tql1}) monotonically increases up
to the point $t=d(d-1)$, whence the replacing is formally valid
under the constraint $\sqrt{d}{\,}\tau\leq{d}(d-1)$, i.e
$\|\rho-\omega\|_{2}\leq2\sqrt{d}(d-1)$. So the upper bound
(\ref{tql2}) is also dealt for all possible values of the
Frobenius norm distance between states (and $0<q<1$). Repeating
the reasons of section 3 of Ref. \cite{rastjst} with new bounds on
the Tsallis entropy, we obtain the following statement.

\begin{Thm}\label{unfn}
Let $\rho$ and $\omega$ be density operators on $d$-dimensional
Hilbert space $\hh$. For the parameter range (\ref{pdm1}) and all
possible values of the distance $\tau=(1/2)\|\rho-\omega\|_{2}$,
there holds
\begin{equation}
\bigl|\rqh(\rho)-\rqh(\omega)\bigr|
\leq\frac{d^{1-q/2}(d-1)^{1-q}\tau^{q}-q\sqrt{d}{\,}\tau}{1-q}
\ . \label{resfk3}
\end{equation}
For the parameter range (\ref{pdp1}) and all possible values of
the distance $\tau=(1/2)\|\rho-\omega\|_{2}$, there holds
\begin{equation}
\bigl|\rqh(\rho)-\rqh(\omega)\bigr|
\leq\varkappa_s{\>}\frac{q\sqrt{d}{\,}\tau-(d-1)^{1-q}d^{q/2}\tau^{q}}{q-1}
\ , \label{reszh3}
\end{equation}
where the factor $\varkappa_s=d^{2(q-1)}$ for $s\in[-1;0]$ and
$\varkappa_s=1$ for $s\in[+1;+\infty)$.
\end{Thm}

Hence we obtain another upper bounds on the map
$(q,s)$-entropies in terms of the quantity (\ref{mp2m}). Designating
$\|\sigma(\Phi)-\sigma(\Psi)\|_{2}=2\tau$, we have
\begin{align}
 & \bigl|\rqm(\Phi)-\rqm(\Psi)\bigr|
\leq\frac{d^{1-q/2}(d-1)^{1-q}\tau^{q}-q\sqrt{d}{\,}\tau}{1-q}
\ , \label{mesfk4}\\
 & \bigl|\rqm(\Phi)-\rqm(\Psi)\bigr|
\leq\varkappa_s{\>}\frac{q\sqrt{d}{\,}\tau-(d-1)^{1-q}d^{q/2}\tau^{q}}{q-1}
\ , \label{meszh4}
\end{align}
in the parameter ranges (\ref{pdm1}) and (\ref{pdp1}),
respectively. In comparison with Eqs. (\ref{mesfk2}) and
(\ref{meszh2}), these bounds are somewhat weaker, but their
validity intervals for $\|\sigma(\Phi)-\sigma(\Psi)\|_{2}$ do not
depend on the dimensionality of the Hilbert space. However, for
sufficiently small values of this distance we prefer the bounds
(\ref{mesfk2}) and (\ref{meszh2}).

\section{Some additivity properties of the map $(q,s)$-entropies}\label{bomap}

In this section, we examine properties of the map
$(q,s)$-entropies with respect to tensor product of a pair of
quantum channels. Broad use of the Tsallis entropy in
non-extensive statistical mechanics stems from the fact that it
does not share the additivity in the following sense \cite{AO01}.
If two random variable $X$ and $Y$ are independent then the
Shannon entropy of the joint distribution
$H_{1}(X,Y)=H_{1}(X)+H_{1}(Y)$. In the quantum regime, the von
Neumann entropy enjoys
${\rm{H}}_{1}(\rho^{Q}\otimes\rho^{R})={\rm{H}}_{1}(\rho^{Q})+{\rm{H}}_{1}(\rho^{R})$.
In general, we have the subadditivity \cite{bengtsson}
\begin{equation}
{\rm{H}}_{1}(\rho^{QR})\leq{\rm{H}}_{1}(\rho^{Q})+{\rm{H}}_{1}(\rho^{R})
\ , \label{prad}
\end{equation}
where reduced densities $\rho^{Q}$ and $\rho^{R}$ are obtained
from $\rho^{QR}$ by taking the partial trace. The equality in Eq.
(\ref{prad}) takes place only for the above case of product
states. The Tsallis entropies do not enjoy the additivity with the
product states, though the subadditivity of quantum $q$-entropy is
still obeyed for $q>1$, namely
\begin{equation}
{\rm{H}}_{q}(\rho^{QR})\leq{\rm{H}}_{q}(\rho^{Q})+{\rm{H}}_{q}(\rho^{R})
\ . \label{trad}
\end{equation}
This fact has been conjectured in Ref. \cite{raggio} and later
proved in Ref. \cite{auden07}. Concerning additivity properties,
the unified entropies succeed to the Tsallis entropies
\cite{hey06}. In particular, we have the subadditivity of the
quantum $(q,s)$-entropy for $q>1$ and $s>1/q$ \cite{rastjst}.
Additivity properties of the map $(q,s)$-entropies with respect to
the tensor product of quantum channels are posed as follows.

\begin{Thm}
Let $\Phi_1$ and $\Phi_2$ be quantum channels. For $q>0$ and all
real $s$, the map $(q,s)$-entropy satisfies
\begin{equation}
\rqm(\Phi_1\otimes\Phi_2)=\rqm(\Phi_1)+\rqm(\Phi_2)+(1-q)s{\,}\rqm(\Phi_1){\,}\rqm(\Phi_2)
\ . \label{padm}
\end{equation}
\end{Thm}

{\bf Proof.} The claim is based on the two point. The first is the
expression for unified $(q,s)$-entropy of a product state, which
is formulated as \cite{hey06}
\begin{equation}
\rqh(\rho\otimes\omega)=\rqh(\rho)+\rqh(\omega)+(1-q)s{\,}\rqh(\rho){\,}\rqh(\omega)
\ . \label{padm1}
\end{equation}
The second is the important result that the dynamical matrix
$\D(\Phi_1\otimes\Phi_2)$ is unitarily similar to the product
matrix $\D(\Phi_1)\otimes\D(\Phi_2)$ \cite{rzf11}, whence
\begin{equation}
\rqm(\Phi_1\otimes\Phi_2)=\rqh\bigl(\sigma(\Phi_1)\otimes\sigma(\Phi_2)\bigr)
\ . \label{padm2}
\end{equation}
Combining Eqs. (\ref{padm1}) and (\ref{padm2}) finally leads to
Eq. (\ref{padm}). $\blacksquare$

For the case $s=0$, the relation (\ref{padm}) gives the additivity
of the R\'{e}nyi map entropies. The latter has been proved in Ref.
\cite{rzf11}. It is known that the rank of the dynamical matrix is
equal to the minimal number of terms needed in the operator-sum
representation \cite{zb04}. So if one of channels $\Phi_1$ and
$\Phi_2$ represents a unitary evolution, then its dynamical matrix
is of rank one and its map $(q,s)$-entropy is zero. Then the
relation (\ref{padm}) claims the additivity of the map
$(q,s)$-entropy. If no one of $\Phi_1$ and $\Phi_2$ represents a
unitary evolution, then both the dynamical matrices are of rank
$\geq2$ and both the map entropies are non-zero. In this case, the
map $(q,s)$-entropy is strictly subadditive for $0<q<1$ and $s<0$
as well as for $1<q$ and $0<s$, i.e.
\begin{equation}
\rqm(\Phi_1\otimes\Phi_2)<\rqm(\Phi_1)+\rqm(\Phi_2)
\ , \qquad\bigl\{0<q<1,{\,}s<0\bigr\}{\bigcup}\bigl\{1<q,{\,}0<s\bigr\}
\ . \label{pdsub}
\end{equation}
Further, the map $(q,s)$-entropy is strictly superadditive for
$0<q<1$ and $0<s$ as well as for $1<q$ and $s<0$, i.e.
\begin{equation}
\rqm(\Phi_1\otimes\Phi_2)>\rqm(\Phi_1)+\rqm(\Phi_2)
\ , \qquad\bigl\{0<q<1,{\,}0<s\bigr\}{\bigcup}\bigl\{1<q,{\,}s<0\bigr\}
\ . \label{pdsup}
\end{equation}
Using the map $(q,s)$-entropies, we can herewith separate quantum
channels that represent a unitary evolution. On the
contrary, the map R\'{e}nyi entropies are additive irrespectively
to channel features \cite{rzf11}.

We shall now extend the inequality of Lindblad \cite{glind91} with
the von Neumann entropy exchange to a certain subclass of unified
entropies. Hence, estimating of the output entropy of maximally
entangled input state will be carried out. This question is arisen
within studies of so-called ''additivity conjecture'' concerning a
product quantum channel \cite{rzf11}. An extension of the Lindblad
inequality is posed as follows.

\begin{Thm}\label{liex}
Let $\rho$ be density operator, and $\Phi$ a quantum channel. For
$q>1$ and $s\geq{q}^{-1}$, the inequality
\begin{equation}
\bigl|\rqh(\rho)-\bah_{q}^{(s)}(\rho,\Phi)\bigr|\leq
\rqh\bigl(\Phi(\rho)\bigr)\leq
\rqh(\rho)+\bah_{q}^{(s)}(\rho,\Phi)
\ , \label{exil}
\end{equation}
including permutations of the three entropies, takes place.
\end{Thm}

{\bf Proof.} Within the proof, we recall meaning of the systems
$E$, $Q$, $R$ from Eqs. (\ref{strqe})--(\ref{qseedf}) and the
respective notation. Using one result of Ref. \cite{auden07}, both
the subadditivity and triangle inequality have been stated for the
quantum $(q,s)$-entropies for $q>1$, $s\geq{q}^{-1}$
\cite{rastjst}. For the reduced densities
\begin{equation}
\rho^{Q'}=\tr_{E}\bigl(\rho^{E'Q'}\bigr)
\ , \qquad
\rho^{E'}=\tr_{Q}\left(\rho^{E'Q'}\right)
\ , \label{rheq}
\end{equation}
obtained from the output density operator $\rho^{E'Q'}$, and the
parameter values $q>1$, $s\geq{q}^{-1}$, we have the relations
\begin{equation}
\bigl|\rqh(\rho^{E'})-\rqh(\rho^{Q'})\bigr|\leq\rqh(\rho^{E'Q'})\leq\rqh(\rho^{E'})+\rqh(\rho^{Q'})
\ . \label{trisq}
\end{equation}
The inequality on the left is the triangle inequality, the
inequality on the right expresses the subadditivity. We then
rewrite (\ref{trisq}) as
\begin{equation}
\left|\bah_{q}^{(s)}(\rho^{Q},\fec)-\rqh\bigl(\fec(\rho^{Q})\bigr)\right|\leq\rqh(\rho^{Q})
\leq\bah_{q}^{(s)}(\rho^{Q},\fec)+\rqh\bigl(\fec(\rho^{Q})\bigr)
\ , \label{trisq1}
\end{equation}
in view of the definition (\ref{qseedf}) and
$\rho^{Q'}=\fec(\rho^{Q})$. It was also used that the systems $Q$,
$R$ initially share the same pure state and the system $R$ itself
is not altered. By doing some simple algebra, we further obtain
\begin{equation}
\left|\rqh\bigl(\fec(\rho^{Q})\bigr)-\rqh(\rho^{Q})\right|\leq\bah_{q}^{(s)}(\rho^{Q},\fec)
\leq\rqh\bigl(\fec(\rho^{Q})\bigr)+\rqh(\rho^{Q})
\ , \label{trisq2}
\end{equation}
and the inequality (\ref{exil}) (in which the superscript $Q$ is
already left out). $\blacksquare$

For the von Neumann entropy, an analog of the inequality
(\ref{exil}) plus permutations was presented by Lindblad
\cite{glind91}. The writers of Ref. \cite{rzf11} used Lindblad's
inequality for estimation of the entropy of an output state
arising from a maximally entangled input state. We shall develop
this issue with respect to the unified $(q,s)$-entropies.

\begin{Thm}\label{mesen}
Let $\Phi_1$ and $\Phi_2$ be quantum channels on $\hh$, and
$|\psi_{+}\rangle\in\hh^{\otimes2}$ a maximally entangled state.
For $q>1$ and $s\geq{q}^{-1}$, there holds
\begin{equation}
\bigl|\rqm(\Phi_1)-\rqm(\Phi_2)\bigr|\leq\rqh\bigl(\Phi_1\otimes\Phi_2(|\psi_{+}\rangle\langle\psi_{+}|)\bigr)
\leq\rqm(\Phi_1)+\rqm(\Phi_2) \ . \label{nesem}
\end{equation}
\end{Thm}

{\bf Proof.} Following the method of Ref. \cite{rzf11}, we
represent the output density operator on the doubled space as
\begin{equation}
\Phi_1\otimes\Phi_2\bigl(|\psi_{+}\rangle\langle\psi_{+}|\bigr)=
(\Phi_1\otimes\id)\circ(\id\otimes\Phi_2)\bigl(|\psi_{+}\rangle\langle\psi_{+}|\bigr)
=\Phi_1\otimes\id\bigl(\sigma(\Phi_2)\bigr)
\ . \label{piip}
\end{equation}
It is easy to check that tracing-out the second space from the
density matrix
$\sigma(\Phi_2)=\id\otimes\Phi_2(|\psi_{+}\rangle\langle\psi_{+}|)$
gives the maximally mixed state $\rho_{*}=\pen/d$ on $\hh$.
Applying Eq. (\ref{exil}) to the density $\sigma(\Phi_2)$ and the
channel $\Phi_1\otimes\id$, we then obtain
\begin{equation}
\bigl|\rqm(\Phi_2)-\bah_{q}^{(s)}(\rho_{*},\Phi_1)\bigr|\leq
\rqh\bigl(\Phi_1\otimes\Phi_2(|\psi_{+}\rangle\langle\psi_{+}|)\bigr)\leq
\rqm(\Phi_2)+\bah_{q}^{(s)}(\rho_{*},\Phi_1)
\ , \label{exil12}
\end{equation}
in view of
$\bah_{q}^{(s)}(\rho_{*},\Phi_1)=\bah_{q}^{(s)}\bigl(\sigma(\Phi_2),\Phi_1\otimes\id\bigr)$.
The definition (\ref{qseedf}) merely gives
$\bah_{q}^{(s)}(\rho_{*},\Phi_1)=\rqh\bigl(\sigma(\Phi_1)\bigr)$,
since any purification of $\rho_{*}$ is a maximally entangled
state. Combining this with (\ref{exil12}) completes the proof.
$\blacksquare$

As in Theorem \ref{liex}, the inequalities with permutations of
the three entropies also hold. Thus, the output entropy of a
maximally entangled input state satisfies the two-sided estimate
(\ref{nesem}) for many values of the parameters $q$ and $s$.
Estimation of such a kind seems to be important in the context of
studies of subadditivity conjecture for a tensor product of two
quantum channels \cite{rzf11}. The proved bounds (\ref{nesem}) are
expressed in terms of the two map $(q,s)$-entropies, which
characterize the decoherent behaviour of involved quantum
channels. For the von Neumann entropy, the inequality
(\ref{nesem}) was derived in Ref. \cite{rzf11}. Some remarks
concerning the quantum Tsallis $q$-entropy are contained therein.
The writers of Ref. \cite{rzf11} also provide another
characterization of a product channel in terms of the minimum
output entropies. We do not consider such entropies here.

\section{Conclusions}\label{concls}

We have discussed some important properties of quantum channels in
terms of the unified $(q,s)$-entropies, which form a family of
two-parametric extensions of the standard Shannon and von Neumann
entropies. In many respects, these entropies are similar to the
standard ones. For given input state, different effects of each
unraveling of a channel result in some probability distribution at
the output. Except for the R\'{e}nyi's $q$-entropies of order
$q>1$, the unified $(q,s)$-entropies of this distribution are all
minimized by the same unraveling of a quantum channel. We have
also specified some class of extremal unraveligs such that their
$(q,s)$-entropies are bounded from below by the quantum
$(q,s)$-entropy of the input state. Several upper bounds of Fannes
type have been derived for the introduced map $(q,s)$-entropies.
The Frobenius norm distance between two rescaled dynamical
matrices is easy to express than the trace norm one. So we have
given new continuity estimates of the unified $(q,s)$-entropies in
terms of the Frobenius norm distance between density operators. If
no of two quantum channels represents a unitary evolution then the
map $(q,s)$-entropy of their tensor product is strictly
subadditive in the range
$\bigl\{0<q<1,{\,}s<0\bigr\}{\bigcup}\bigl\{1<q,{\,}0<s\bigr\}$
and strictly superadditive in the range
$\bigl\{0<q<1,{\,}0<s\bigr\}{\bigcup}\bigl\{1<q,{\,}s<0\bigr\}$.
Extending Lindblad's inequality, we have obtained a two-sided
estimate on the output $(q,s)$-entropy for the tensor product of
two channels acting on maximally entangled input state. Overall,
the map entropies of considered kind enjoy useful properties and
may be applied in the context of quantum information processing.

\appendix

\section{Some relations between Schatten norms}\label{srsn}

In this appendix, we examine relations between different Schatten
norms of the same operator. Few results of such a kind were
presented in section IV of Ref. \cite{AE05}. The statement of
Lemma 3 therein allows to give upper bound on a unitarily
invariant norm $|||\ax|||$ of traceless Hermitian $\ax$ in terms
of the trace norm $\|\ax\|_{1}$. In particular, we have
$\sqrt{2}{\>}\|\ax\|_{2}\leq\|\ax\|_{1}$ for all traceless
Hermitian $\ax\in{\mathcal{L}}(\hh)$ \cite{AE05}. In the present
work, however, we are rather needed in upper bound on the trace
norm in terms of the Frobenius one. In general, the following
statement takes place.

\begin{Lem}\label{q12b}
Let $\ax$ be an operator with $d$-dimensional support. For $p,q\geq1$, there holds
\begin{equation}
\|\ax\|_{p}\leq{d}^{(q-1)/(pq)}{\>}\|\ax\|_{pq}
\ , \label{npnpq}
\end{equation}
with the equality if and only if the $\ax$ acts on its support as a multiple of unitary operator.
\end{Lem}

{\bf Proof.} We can restrict our consideration to the support of
$\ax$, in which its singular values are non-zero. In line with the
H\"{o}lder inequality, we have \cite{hardy}
\begin{equation}
|\langle{\rm{u}}{\,},{\rm{v}}\rangle|
\leq\|{\rm{u}}\|_{q}{\>}\|{\rm{v}}\|_{r}
\ , \label{hlin}
\end{equation}
where the conjugate indices $q$ and $r$ obey $1/q+1/r=1$ and the vector norms are
\begin{equation}
\|{\rm{u}}\|_{q}=\left(\sum\nolimits_{j=1}^{d}{|u_{j}|^{q}}\right)^{1/q}
\ , \qquad
\|{\rm{v}}\|_{r}=\left(\sum\nolimits_{j=1}^{d}{|v_{j}|^{r}}\right)^{1/r}
\ . \label{vecnom}
\end{equation}
Putting $u_{j}=\vrs_j(\ax)^{p}$ and $v_{j}=1$ for all $j$, we then get
\begin{equation}
\sum\nolimits_{j=1}^{d}{\vrs_j(\ax)^{p}}\leq
\left(\sum\nolimits_{j=1}^{d}{\vrs_j(\ax)^{pq}}\right)^{1/q}d^{1-1/q}
\ . \label{gear}
\end{equation}
Raising both the sides to the power $1/p$, we finally obtain
(\ref{npnpq}). The equality in (\ref{hlin}) takes place if and
only if the $d$-tuples $\rm{u}$ and $\rm{v}$ are linearly related
(see, e.g., theorem 14 in \cite{hardy}). Hence the equality in
(\ref{gear}) is equivalent to that $\vrs_j=\vrs_k$ for all
$j\neq{k}$. In the basis of its eigenstates, the $|\ax|$ is then
rewritten as $|\ax|=\vrs_1\pen$, whence the operator
$\vrs_1^{-1}\ax$ is unitary. $\blacksquare$

The relation (\ref{npnpq}) gives an upper bound on the Schatten
$p$-norm in terms of other Schatten norms with larger values of
the parameter. Recall that the Schatten $p$-norm is non-increasing
in $p$ \cite{watrous1}. In particular, for $p=1$ we have an
estimate on the trace norm expressed as
\begin{equation}
\|\ax\|_{1}\leq{d}^{(q-1)/q}{\>}\|\ax\|_{q}
\ . \label{n1npq}
\end{equation}
Note that the inequality (\ref{npnpq}) remains valid for
$q\to\infty$ and finite $p$, namely
\begin{equation}
\|\ax\|_{p}\leq{d}^{1/p}{\>}\|\ax\|_{\infty}
\ . \label{ninpq}
\end{equation}
This is an upper bound on the Schatten $p$-norm in terms of the
spectral one. Conditions for the equality in (\ref{ninpq}) are the
same as for the equality in (\ref{npnpq}).

\end{document}